\documentclass[10pt,twocolumn,letterpaper]{article}
\usepackage[pagenumbers]{cvpr} 

\usepackage{graphicx}
\usepackage{amsmath}
\usepackage{amssymb}
\usepackage{booktabs}
\usepackage[accsupp]{axessibility}
\makeatletter
\@namedef{ver@everyshi.sty}{}
\makeatother
\usepackage{cuted}

\usepackage[pagebackref,breaklinks,colorlinks]{hyperref}

\usepackage[capitalize]{cleveref}
\crefname{section}{Sec.}{Secs.}
\Crefname{section}{Section}{Sections}
\Crefname{table}{Table}{Tables}
\crefname{table}{Tab.}{Tabs.}

\def\cvprPaperID{6} 
\def\confName{CVMIworkshop}
\def\confYear{2022}

\makeatletter
\@namedef{ver@everyshi.sty}{}
\makeatother
\usepackage{changes}
\definechangesauthor[name={Per cusse}, color=orange]{per}
\usepackage{lipsum}
\newcommand\blfootnote[1]{%
\begingroup
\renewcommand\thefootnote{}\footnote{#1}%
\addtocounter{footnote}{-1}%
\endgroup
}
\begin{document}

\title{BCI: Breast Cancer Immunohistochemical Image \\ Generation through Pyramid Pix2pix}
\author{
Shengjie Liu\textsuperscript{\rm 1} \space \space
Chuang Zhu\thefootnote{$^*$}\textsuperscript{\rm 1 } \space \space
Feng Xu\thefootnote{$^*$}\textsuperscript{\rm 2 } \space \space
Xinyu Jia\textsuperscript{\rm 1} \space \space
Zhongyue Shi\textsuperscript{\rm 2} \space \space
Mulan Jin\textsuperscript{\rm 2} \space \space\\
\textsuperscript{\rm 1}Beijing University of Posts and Telecommunications, Beijing, China\\
\textsuperscript{\rm 2}Capital Medical University, Beijing, China\\
{\tt\small \{shengjie.Liu, czhu, jiaxinyubupt\}@bupt.edu.cn}\\
{\tt\small drxufeng@mail.ccmu.edu.cn}\space \space \space
{\tt\small \{shizhongyue815, kinmokuran\}@163.com}\space \space \space \\
 }

 \maketitle

\begin{strip}
    \centering
    \vspace{-40pt}
    \includegraphics[trim={0cm 0cm 0cm 0cm},clip,width=\textwidth]{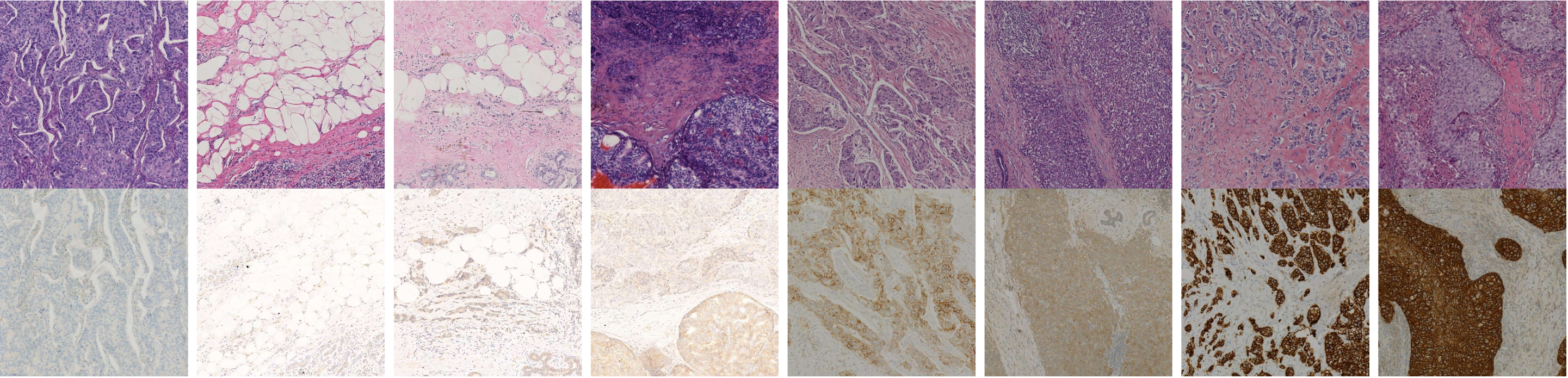}
    \vspace{-2.em}
    \captionof{figure}{Samples of BCI. \textbf{Top}: HE-stained patches. \textbf{Bottom}: IHC-stained patches. Each column represents a HE-IHC image pair. It contains four expression levels of HER2 (0, 1+, 2+, 3+).}
    \label{samples}
\end{strip}
\begin{abstract}
The evaluation of human epidermal growth factor receptor 2 (HER2) expression is essential to formulate a precise treatment for breast cancer. The routine evaluation of HER2 is conducted with immunohistochemical techniques (IHC), which is very expensive. Therefore, for the first time, we propose a breast cancer immunohistochemical (BCI) benchmark attempting to synthesize IHC data directly with the paired hematoxylin and eosin (HE) stained images. The dataset contains 4870 registered image pairs, covering a variety of HER2 expression levels. 
  
Based on BCI, as a minor contribution, we further build a pyramid pix2pix image generation method, which achieves better HE to IHC translation results than the other current popular algorithms. Extensive experiments demonstrate that BCI poses new challenges to the existing image translation research. Besides, BCI also opens the door for future pathology studies in HER2 expression evaluation based on the synthesized IHC images. BCI dataset can be downloaded from \url{https://bupt-ai-cz.github.io/BCI}.
\end{abstract}
\blfootnote{*Corresponding authors: Chuang Zhu (czhu@bupt.edu.cn), Feng Xu (drxufeng@mail.ccmu.edu.cn)}
\section{Introduction}
\label{sec:intro}

\begin{figure}[hbt]
  \centering
  \begin{subfigure}{0.48\linewidth}
    \includegraphics[width=1\linewidth]{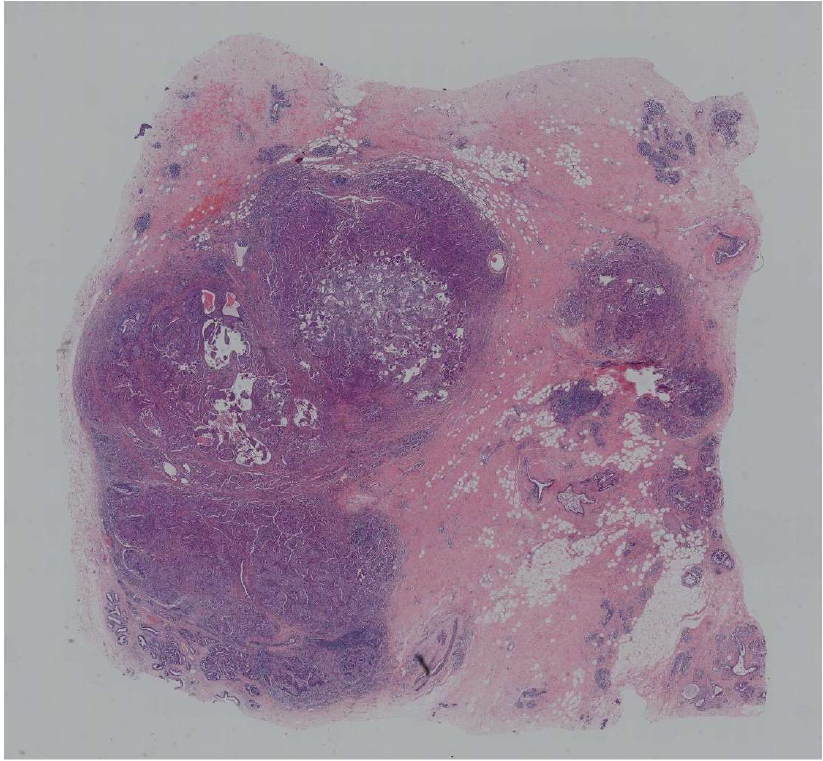}
    \caption{An example of HE slice.}
    \label{HEexample}
  \end{subfigure}
  \hfill
  \begin{subfigure}{0.48\linewidth}
    \includegraphics[width=1\linewidth]{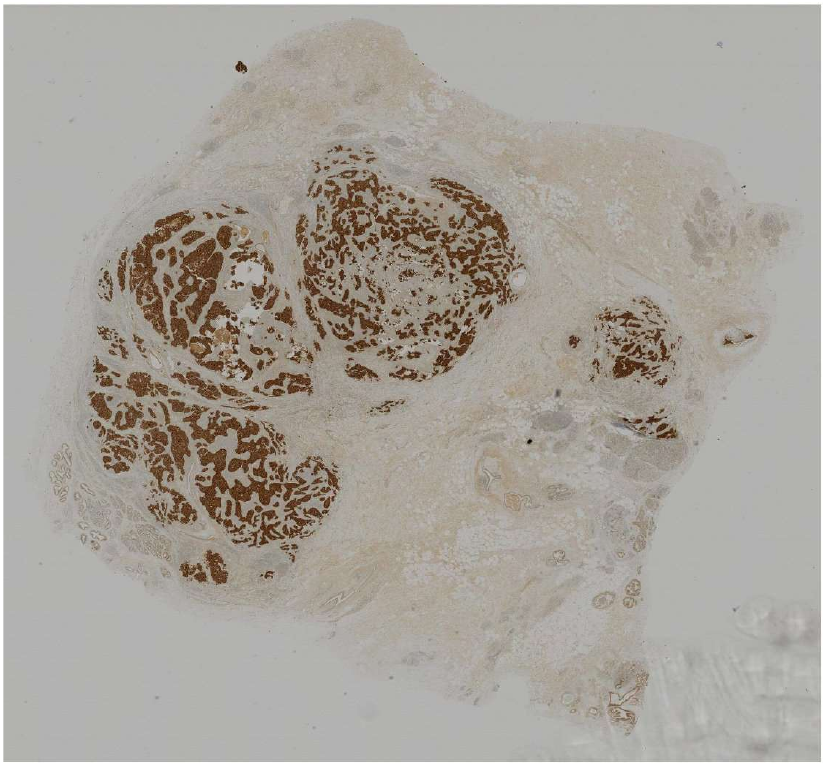}
    \caption{An example of IHC slice.}
    \label{HER2example}
  \end{subfigure}
  \caption{Visualization of HE-stained and IHC-stained slices.}
  \label{WSIexamples}
\end{figure}

According to work \cite{torre2017global}, breast cancer is a leading cause of death for women. Accurate diagnosis and therapy are key factors to reduce the mortality rate of breast cancer patient \cite{zhu2019breast}. The histopathological checking is a gold standard to identify breast cancer. To achieve this, the tumor materials are first made into hematoxylin and eosin (HE) stained slices (a slice is shown in Fig.~\ref{WSIexamples}(a)). Then, the diagnosis is performed by pathologists through observing the HE slices under the microscope or analyzing the digitized whole slice images (WSI). For diagnosed breast cancer, it is essential to formulate a precise treatment plan by checking the expression of specific proteins, such as human epidermal growth factor receptor 2 (HER2) \cite{la2020detection}. The breast cancer with over-expression of HER2 is prone to have aggressive clinical behaviour, and thus accurate therapy should be formulated accordingly. 

\begin{figure}[hbt]
  \centering
  \begin{subfigure}{0.48\linewidth}
    \includegraphics[width=1\linewidth]{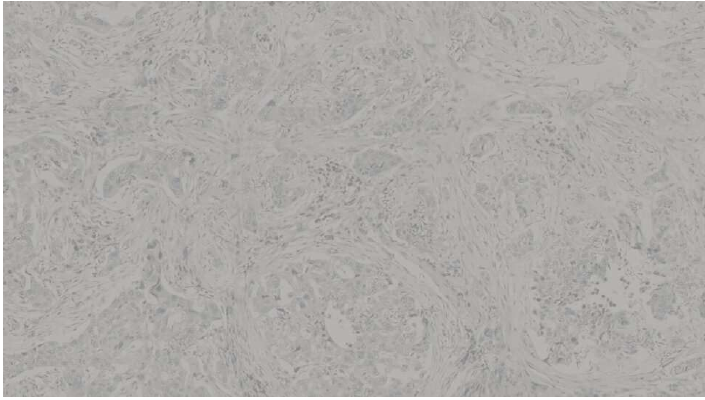}
    \caption{IHC 0}
    \label{IHC(0)}
  \end{subfigure}
  \hfill
  \begin{subfigure}{0.48\linewidth}
    \includegraphics[width=1\linewidth]{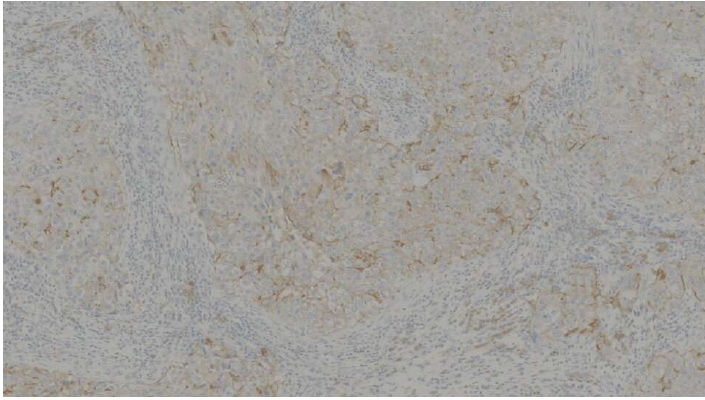}
    \caption{IHC 1+}
    \label{IHC(1+)}
  \end{subfigure}
  \hfill
  \begin{subfigure}{0.48\linewidth}
    \includegraphics[width=1\linewidth]{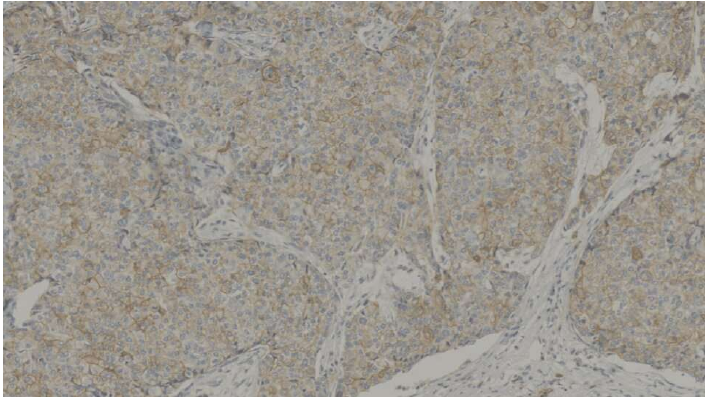}
    \caption{IHC 2+}
    \label{IHC(2+)}
  \end{subfigure}
  \hfill
  \begin{subfigure}{0.48\linewidth}
    \includegraphics[width=1\linewidth]{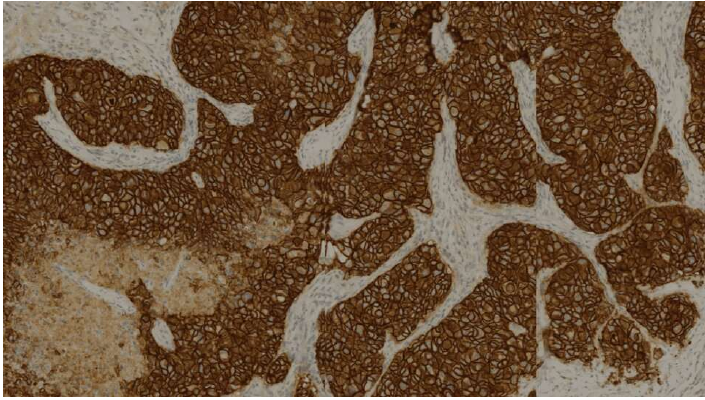}
    \caption{IHC 3+}
    \label{IHC(3+)}
  \end{subfigure}
  \caption{Visualization of four kinds of HER2 expressions.}
  \label{HER2_expression}
\end{figure}

The routine evaluation of HER2 expression is conducted with immunohistochemical techniques (IHC) \cite{khameneh2019automated}. Specifically, one additional IHC-stained slice (a slice is shown in Fig.~\ref{WSIexamples}(b)) is first prepared. Then the pathologists will check the IHC-stained slice to obtain the HER2 expression status: IHC 0, no staining is observed or membrane staining that is incomplete and is faint/barely perceptible and in $\leq 10\%$ of tumor cells (Fig.~\ref{HER2_expression}(a)); IHC 1+, incomplete membrane staining that is faint/barely perceptible and in \textgreater 10\% of tumor cells (Fig.~\ref{HER2_expression}(b)); IHC 2+, weak to moderate complete membrane staining observed in \textgreater 10\% of tumor cells (Fig.~\ref{HER2_expression}(c)); IHC 3+, circumferential membrane staining that is complete, intense, and in \textgreater 10\% of tumor cells (Fig.~\ref{HER2_expression}(d)) \cite{wolff2018human}. The detection of HER2 expression is critical to the formulation of follow-up treatment plans for breast cancer. However, it is very expensive to conduct HER2 evaluation through the additional preparation of IHC-stained slice. Then the question is can we synthesize the IHC-stained image based on HE-stained WSI? In case of success, we can conduct HER2 expression evaluation directly based on the synthesized IHC-stained slices.

This paper presents the above challenge for the first time, and tries to solve it through image-to-image translation technique. Image translation aims to learn the mapping between an input source-domain image and an output target-domain image \cite{huang2020multimodal}. In recent years, some methods and datasets have been proposed to promote the research of image-to-image translation.

Pix2pix\cite{isola2017image} proposes a universal translation method for paired images. Since then, there have been other supervised image translation algorithms based on pix2pix that can be applied to specific scenes: pix2pixHD\cite{wang2018high} has achieved very good results in high resolution paired image translation; work \cite{qu2019enhanced} proposes an enhanced pix2pix optimized for image dehazing. Besides, there are also many excellent methods\cite{liu2017unsupervised,lee2020drit++,huang2018multimodal,chong2021gans} for unsupervised image translation inspired by these pioneering works \cite{zhu2017unpaired,kim2017learning,yi2017dualgan}.

Dataset is the key factor for image translation, especially for supervised methods. Many fields have proposed datasets pertinently. However, there are only a few works for the medical image translation applications. RegGAN\cite{kong2021breaking} implements a general image translation method for both paired and unpaired images on BraTS\cite{menze2014multimodal} dataset. In the field of breast cancer, a few of datasets such as BCNB \cite{xu2021predicting} have been proposed for automatic diagnosing, however, there are no datasets for HE to IHC staining for HER2 detection. This task requires structural level aligned datasets, which poses great challenges due to the difficulty of the acquisition of well paired HE-IHC images. To the best of our knowledge, there are no public image translation datasets exists for HER2 detection in breast cancer tissue.

To spur research in this area, we introduce BCI, a structural aligned dataset for the translation of HE-stained slices to immunohistochemical results (Fig.~\ref{samples}). We also propose a method optimized for this task. We benchmark several state-of-the-art (SOTA) algorithms for image translation tasks. In summary, this paper makes the following contributions:

\begin{itemize} 
\item We collect and build BCI: a paired HE to HER2 expression image translation dataset. To our knowledge, BCI is the first large-scale publicly available dataset for immunohistochemical image generation.

\item We propose a pyramid pix2pix method to generate immunohistochemical image based on HE. Compared with other pix2pix-like methods\cite{isola2017image,wang2018high}, our method can constrain the generated image at multiple scales and achieve better results on our BCI dataset. 

\item We conduct extensive experiments on BCI and LLVIP dataset\cite{jia2021llvip} to explore the gains that different scales bring to the model, which demonstrate the flexibility and versatility of multi-scale constraints. 

 \end{itemize}

\begin{figure*}[htb]
\begin{center}
\includegraphics[width=0.98\linewidth]{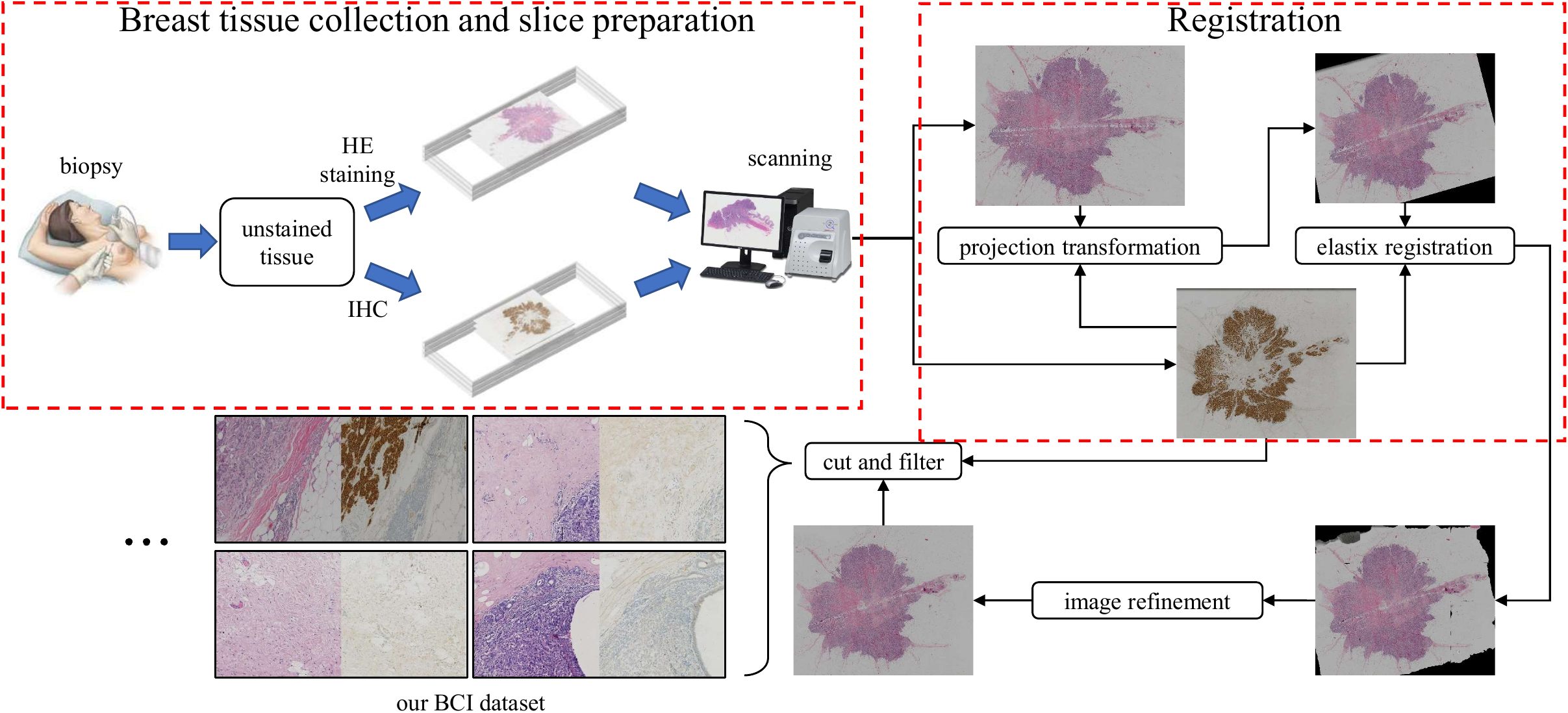}
\end{center}
\caption{The establishment of our BCI dataset is generally divided into three steps: 1) breast tissue collection and slice preparation; 2) registration of images in the two domains; 3) post-processing including image refinement and patches cutting.} 
\label{imageprocess} 
\end{figure*}

\section{Related Work}
\label{sec:formatting}

In this section, we will review some image translation algorithms, as well as some datasets that are often used in image translation tasks.

\subsection{Image Translation}
Image translation algorithm establishes a mapping between two domain images. It is often used in image semantic, image synthesis, and image super-resolution, etc. Image translation algorithms can be divided into unsupervised image translation and supervised image translation.

Unsupervised image translation does not require aligned datasets, which makes the application range of this method very wide. Many works \cite{zhu2017unpaired,kim2017learning,yi2017dualgan,lee2020drit++,liu2017unsupervised,huang2018multimodal,chong2021gans} are dedicated to the translation of unpaired images. This type of methods can randomly extract images from two domains during training, which can achieve better image style transfer when it is difficult to obtain paired data. In addition to solving the problem of image translation from one domain to another, there are some methods \cite{choi2018stargan,choi2020stargan} that creatively solve the image translation between multiple domains.

However, unsupervised methods also have a certain limitation: for paired datasets, it may not be possible to establish an accurate mapping between the two domains. To overcome this problem, work \cite{xu2019gan} adds additional per-patch labels (e.g. background, necrosis, fibrosis, etc.) to CycleGAN during training. For the classification of patches, huge amounts of human work are still required. Therefore, supervised image translation methods still have great application value. Pix2pix\cite{isola2017image} is a pioneering supervised image translation algorithm. It is a general image translation algorithm that can be applied to various image translation tasks. In addition to the adversarial loss between the generator and the discriminator, it also calculates the pixel-level difference between the generated image and the ground truth to continuously improve the generator's effect. Pix2pixHD\cite{wang2018high} optimizes the generator structure on the basis of pix2pix to make the generation of high-resolution images better. EPDN\cite{qu2019enhanced} follows the overall structure of pix2pix, which uses a multi-resolution generator, a multi-scale discriminator, and an enhancer for image dehazing tasks. SRGAN\cite{ledig2017photo} and ESRGAN\cite{wang2018esrgan} apply the generative adversarial network to image super-resolution tasks, which are essentially a kind of image translation. There are also some supervised methods\cite{park2019semantic,sushko2020you} that are widely used in semantic image synthesis. These models take semantic information as input and translate it into real images. 

In the medical field, image translation already has some applications. RegGAN\cite{kong2021breaking} proposes a general image translation model, which adds a U-net structure registration network after the generator in pix2pix. It calculates the loss between the output image of the registration network and the ground truth, which makes RegGAN achieve good results in both paired and unpaired data. In the field of pathological images, there are some works that \cite{cho2017neural,shaban2019staingan} translate non-standard stained sections into standard stained sections, providing new ideas for the normalization of pathological image staining.

\subsection{Datesets}

The datasets used for image translation tasks are abundant, and these datasets can also be divided into two categories: paired and unpaired. Many paired datasets \cite{Tylecek13,richter2016playing,cordts2016cityscapes,zhou2017scene,caesar2018coco,ros2016synthia} can be used for translation between semantic distribution maps and real images; Cityscapes{\cite{cordts2016cityscapes} and Foggy Cityscapes\cite{sakaridis2018semantic} can be used for image dehazing research; CelebAMask-HQ\cite{lee2020maskgan} and FFHQ-Aging\cite{or2020lifespan} are two large-scale face datasets}, are used in the research of generating face images from segmentation masks. Part of the images in work\cite{laffont2014transient} can be used for conversion between day and night. LLVIP\cite{jia2021llvip} contains registered images in two domains of visible light and infrared light, which can be used for translation between visible light images and infrared light images. BraTS\cite{menze2014multimodal} is a dataset in the medical field, in which T1 weighted images and T2 weighted images can be used for the translation of brain MRI images. Selfie2anime\cite{kim2019u} is an unpaired dataset that provides images in two domains, selfies and cartoon characters, which can be used for the research of transforming real pictures into cartoon styles. There are also some multi-domain datasets such as AFHQ\cite{choi2020stargan} and RaFD\cite{langner2010presentation}, these datasets can be used for unsupervised image translation and image synthesis in multiple domains. Note that all paired datasets can be used to train unsupervised image translation models.

\section{BCI Dataset}

The application of deep learning in the medical field has attracted more and more attention. There are already some brain image translation datasets to promote research on brain science, however, there is still no relevant data for pathological image translation. Therefore, we propose the BCI dataset, in order to better promote the research of pathological image translation. We hope that BCI can play a positive role in the diagnosis of breast cancer. At the same time, as a benchmark, our dataset can help analyze the advantages and disadvantages of current image translation algorithms.

Our overall process of building the dataset is shown in Fig.~\ref{imageprocess}. Next, we will introduce more details about this dataset.

\subsection{Collection}

The data scanning equipment is Hamamatsu NanoZommer S60, a pathology section scanner with a scanning speed of 60 seconds per slice. The scanning resolution of the equipment is 0.46 $\mu m$ per pixel. We scanned more than 600 pathological slices of breast cancer tissues and sorted out the WSI stained with HE and the corresponding immunohistochemical WSI of 319 breast cancer patients. In the image registration process, we filter out WSI pairs that are unable to complete the alignment. Finally, we got 4870 pairs of HE-IHC patches from 51 different WSI image pairs.

\subsection{Registration}

For a piece of pathological tissue, the doctor will cut two tissue samples from it for HE staining and HER2 detection. Therefore, there will be differences in the morphology of the two pathological samples. Besides, the tissue samples will be stretched or squeezed to a certain extent during slice preparation, which will increase the difference between the samples. In order to make the images of the two domains aligned, we need to perform registration processing on the images.
\begin{figure}[htb]
\begin{center}
\includegraphics[height=0.46\linewidth]{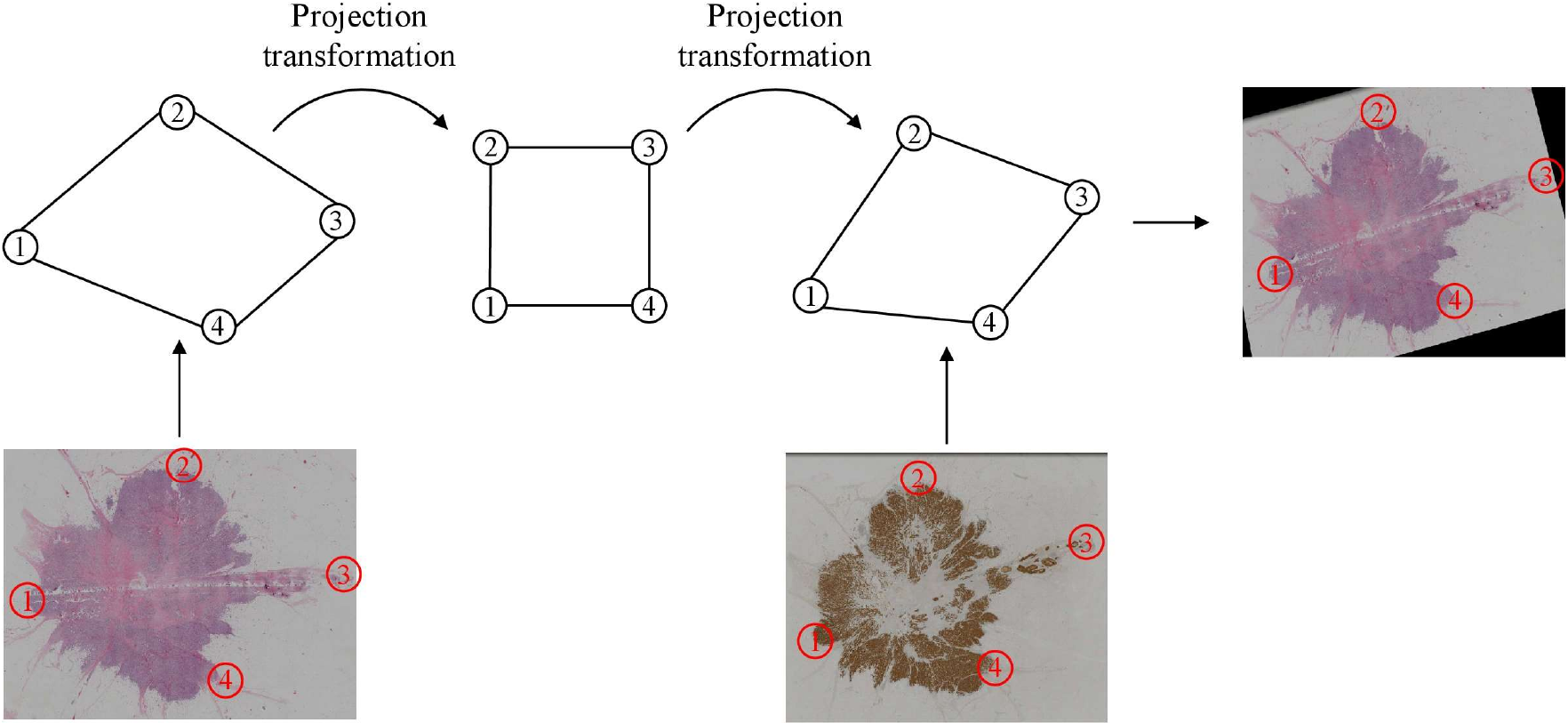}
\end{center}
\caption{Projection transformation by manually selecting corresponding points. The HE image can be initially aligned with the IHC image after a two-step projection transformation.} 
\label{projection_transformation} 
\end{figure}

\noindent{\textbf{Projection transformation.}} First, we use the method of human-computer interaction projection transformation to roughly align the WSI pairs. This method requires manually selecting no less than 4 pairs of corresponding points on the two WSIs (Fig.~\ref{projection_transformation}), then through the method of projection mapping, the irregular quadrilateral determined by the four points on the HE WSI is first mapped to a square, and then the square is mapped to the irregular quadrilateral determined by the four points on the IHC WSI. In this process, the HE image is basically aligned with the contour of the corresponding IHC image by translation and rotation, and the resolution of the two images is kept consistent. At this time, there are still some deviations inside the HE and IHC images, and further registration is required. 
\begin{figure}[htb]
  \centering
   \includegraphics[width=0.98\linewidth]{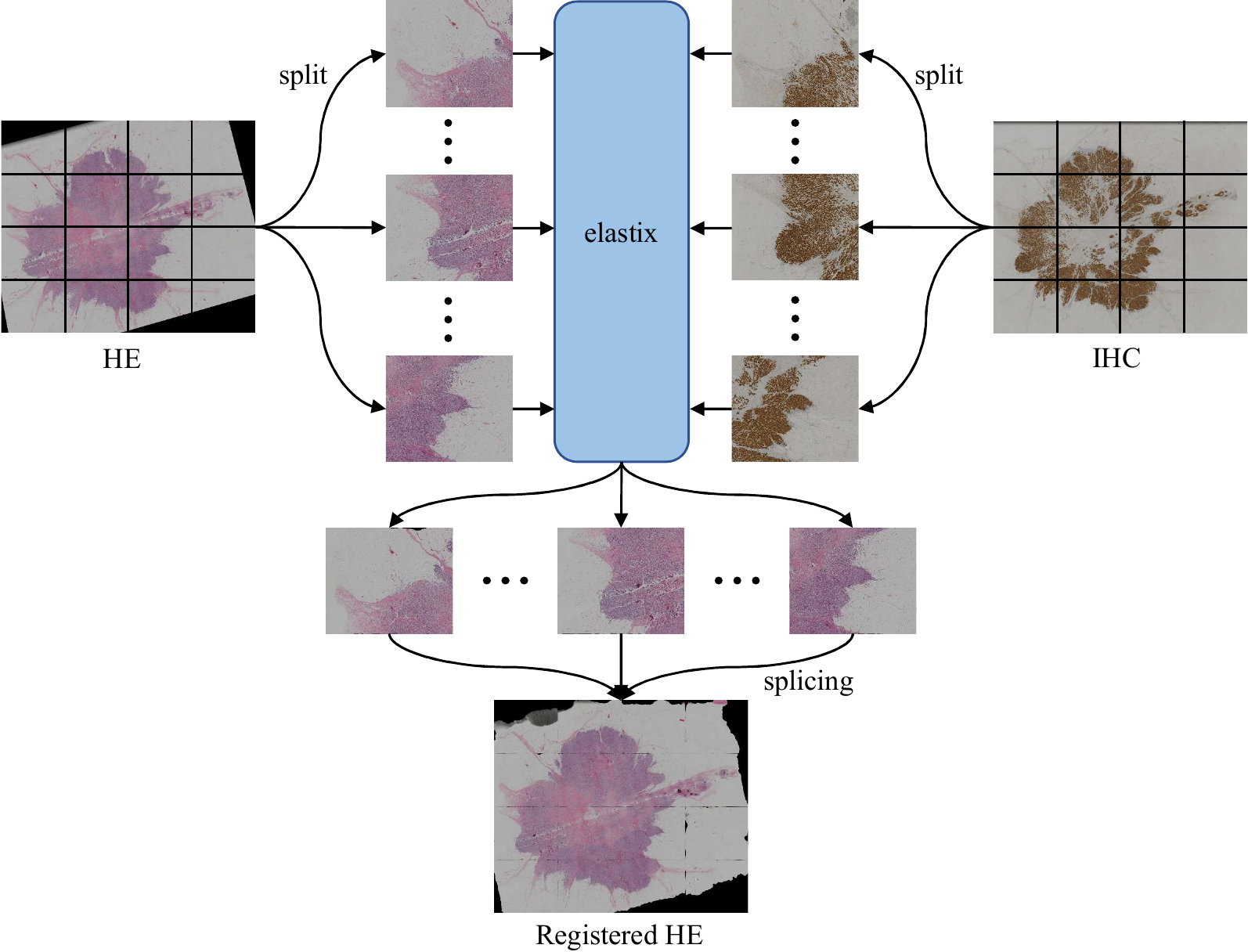}
   \caption{The process of elastix registration. The roughly aligned HE and IHC images are divided into blocks, and each block is registered separately with elastix. Finally, the registered blocks are re-spliced}
   \label{elastix}
\end{figure}

\begin{figure}[htb]
\begin{center}
\includegraphics[height=0.49\linewidth]{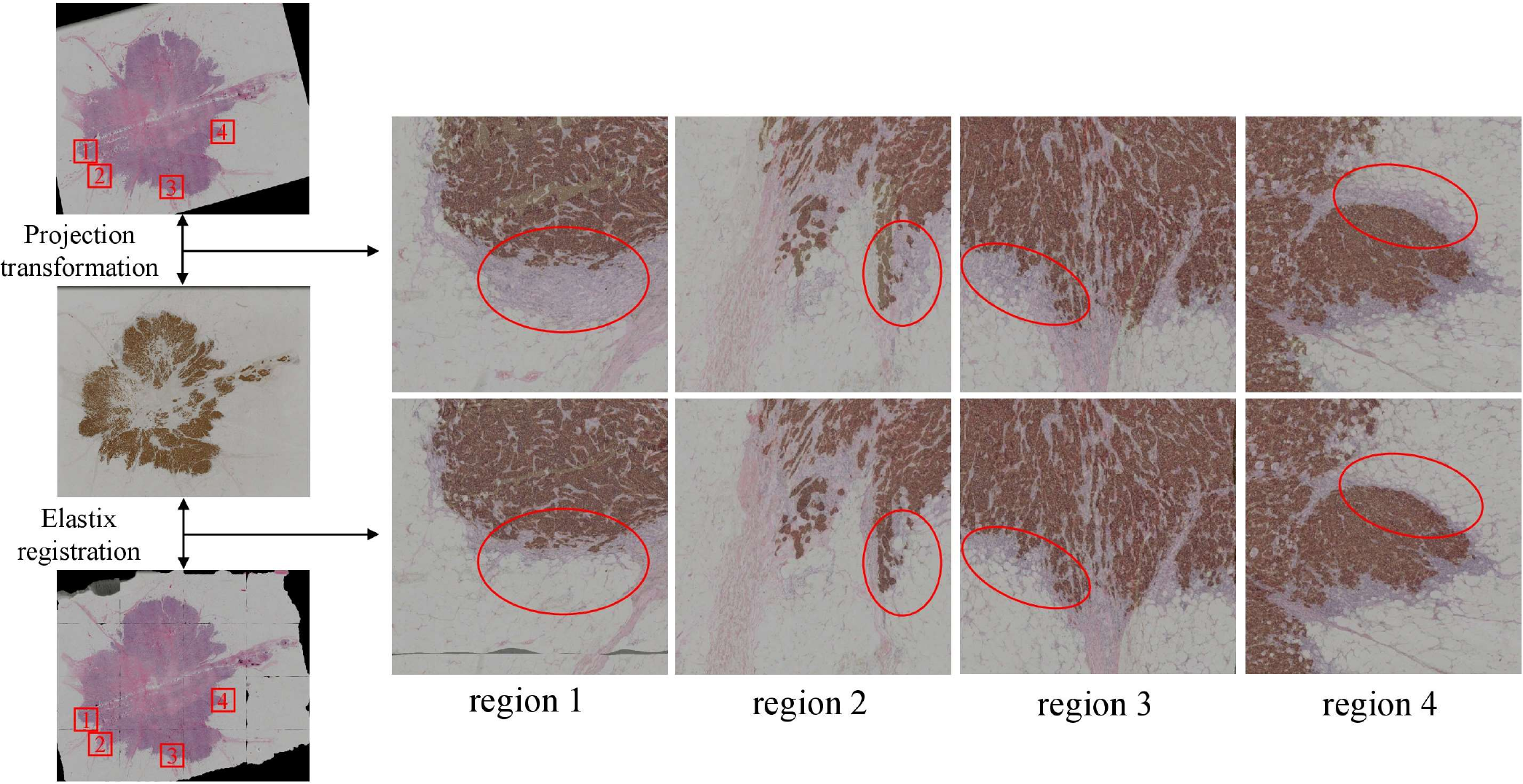}
\end{center}
\caption{By overlapping the registration result with the corresponding IHC image, the difference between projection transformation and elastix registration can be seen: projection transformation can only roughly overlap the two images, but cannot achieve detailed registration; after elastix registration, the overlap in details is realized.} 
\label{registration} 
\end{figure}

\begin{figure}
  \centering
    \includegraphics[width=0.98\linewidth]{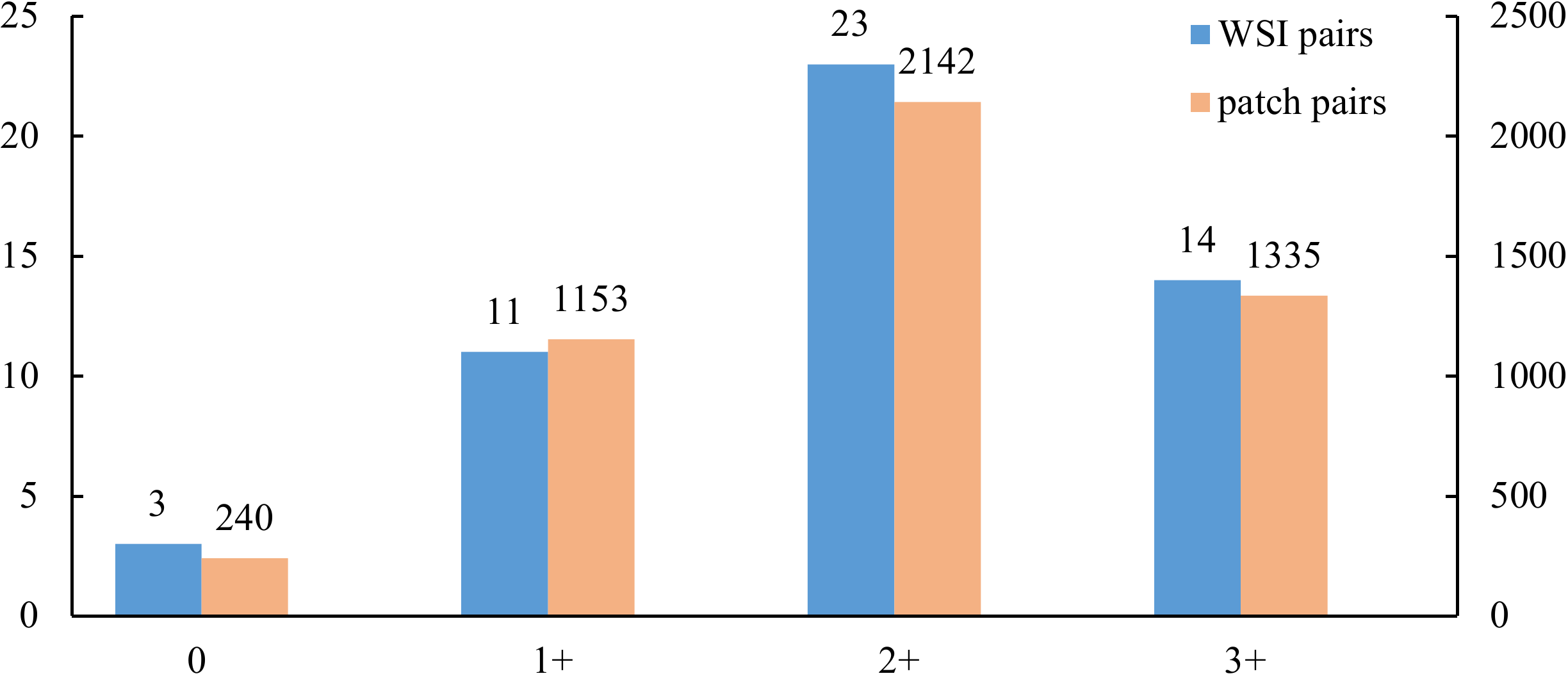}
  \caption{Image statistics of IHC results.}
  \label{tumor_statistics}
\end{figure}

\begin{figure*}[t]
  \centering
   \includegraphics[width=0.98\linewidth]{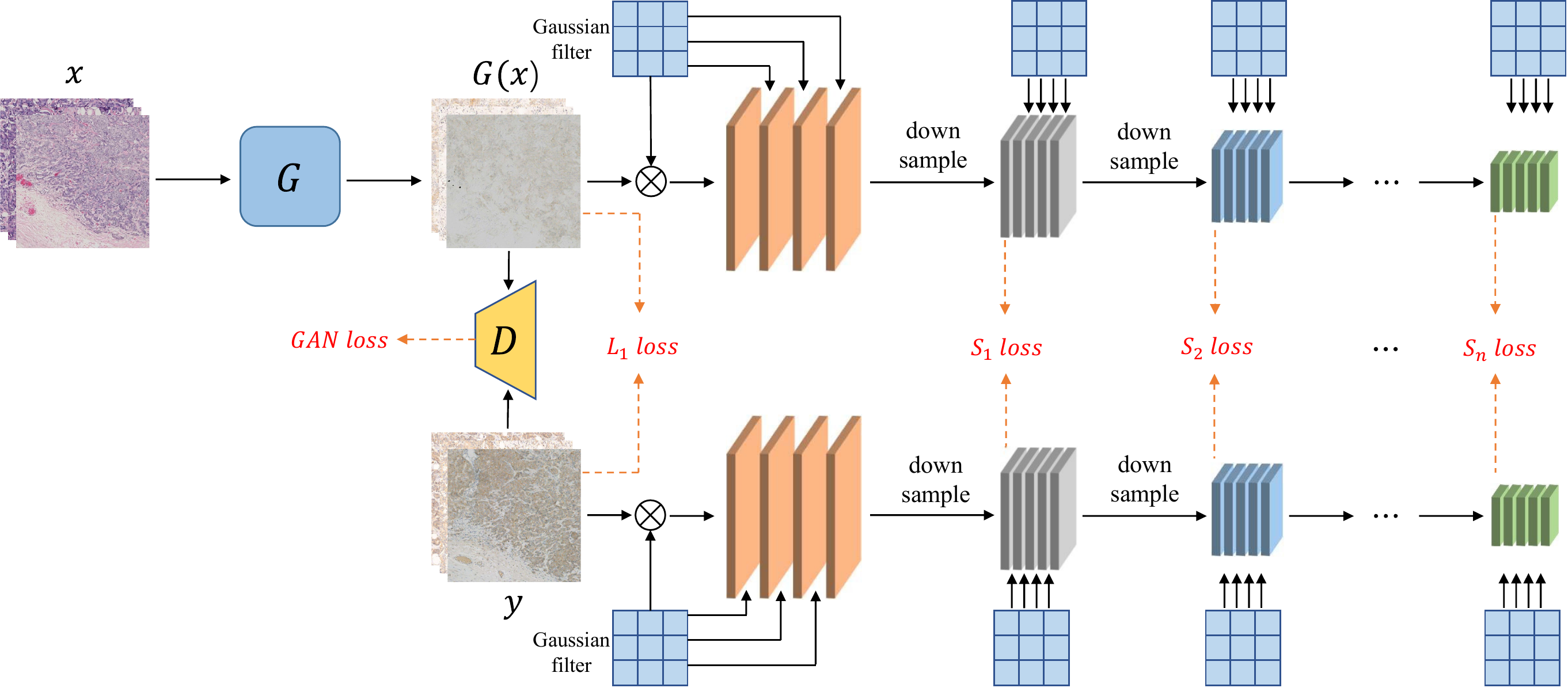}
   \caption{The framework of the proposed pyramid pix2pix. The image of each scale is obtained from the image of the previous scale after four Gaussian convolutions and one downsampling.}
   \label{overallstructure}
\end{figure*}

\noindent{\textbf{Elastix registration.}} Second, we use the registration toolbox elastix\cite{klein2009elastix} to perform fine-grained regional non-rigid registration. This process can align the details of the two domain images as much as possible. Fig.~\ref{registration} shows the detailed alignment of elastix registration based on projection transformation. For each WSI, because its resolution is too high (about 20,000 pixels on a side), the computational power and time consumed by direct registration are huge, in order to improve the efficiency of registration, we divide it into 16 blocks for registration respectively. At the same time, since elastix cannot directly process the RGB image, we split the image and use only a single channel for registration, save the transformation file of the channel, and then apply the transformation file to the other two channels. Finally, we merge the registered images of the three channels and then stitch each registered block into WSI. The registration process of elastix is shown in Fig.~\ref{elastix}.

\subsection{Post-processing}

Our block registration method is efficient, however, during the registration process, the expansion and contraction of the image will leave a gap on the edge of each image block. Therefore, we need to remove the black border between the blocks and fill it with the surrounding content. Finally, the registered WSI image is cut into 1024$\times$1024 size patches. Finally, we will filter out blank and not well-aligned areas. 

Our BCI dataset contains 4870 pairs of pediatric pathological image patches with a resolution of 1024$\times$1024. These patches are from the WSIs of 51 patients. The immunohistochemical results of these 51 patients included four categories: 0, 1+, 2+, and 3+. Fig.~\ref{tumor_statistics} shows the distribution of the 51 WSIs and the number of patches from different IHC results.

\section{Proposed Method}
\subsection{Architecture}

Our BCI dataset presents a new challenge for image translation. In our dataset, the images of the two domains are paired and registered at the structural level. However, due to the existence of image differences between the two domains, some positions cannot achieve pixel-level alignment, which makes the existing pix2pix series of algorithms difficult to work; at the same time, we need to perform targeted output for each HE stained image, which is not the strength of the unsupervised algorithms. Therefore, we propose a pyramid pix2pix model suitable for structural aligned data. Our overall framework is shown in Fig.~\ref{overallstructure}.

\begin{figure}
  \centering
  \begin{subfigure}{0.24\linewidth}
    \includegraphics[width=1\linewidth]{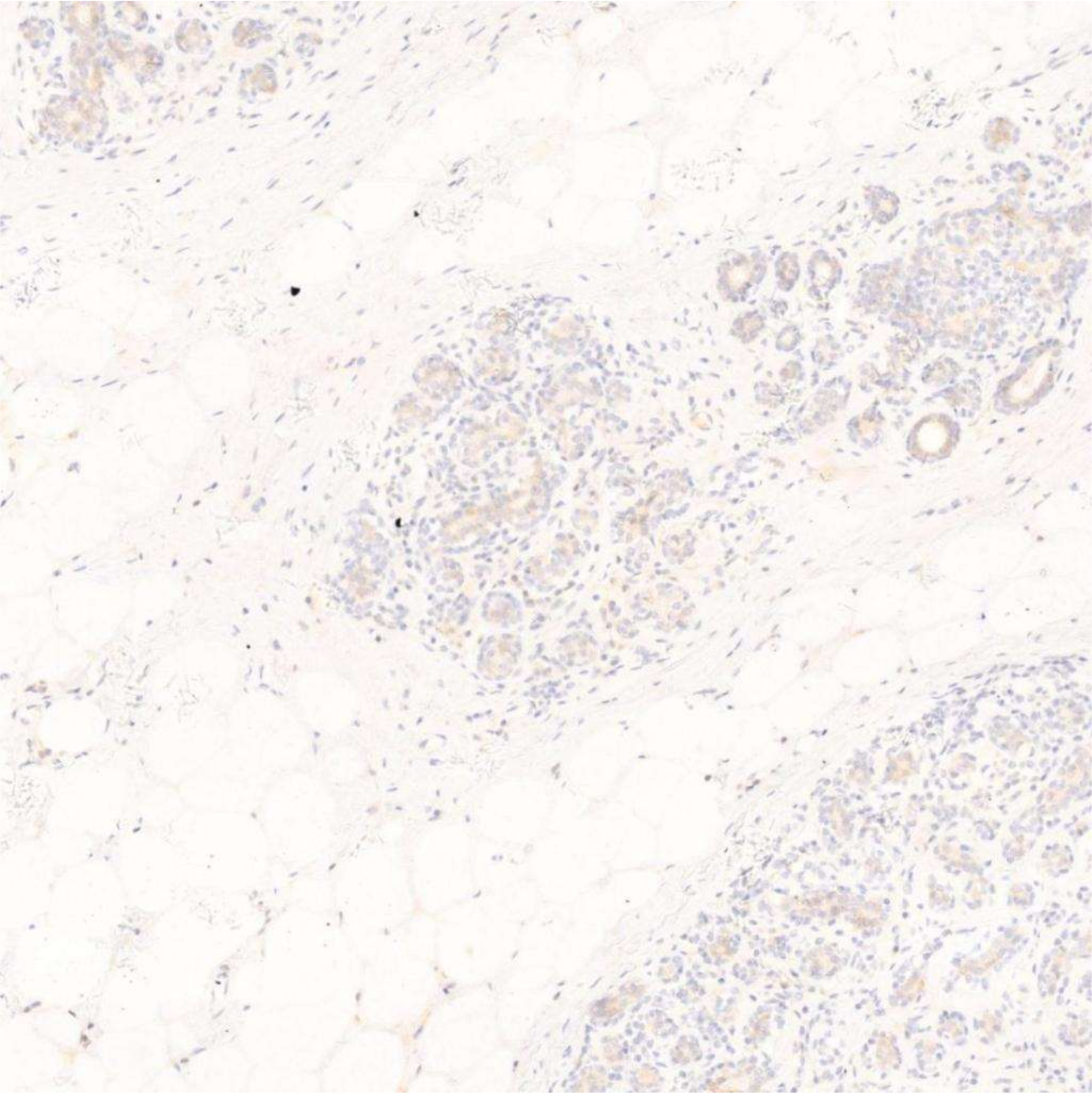}
    \caption{scale 1}
    \label{Octave1}
  \end{subfigure}
  \hfill
  \begin{subfigure}{0.24\linewidth}
    \includegraphics[width=1\linewidth]{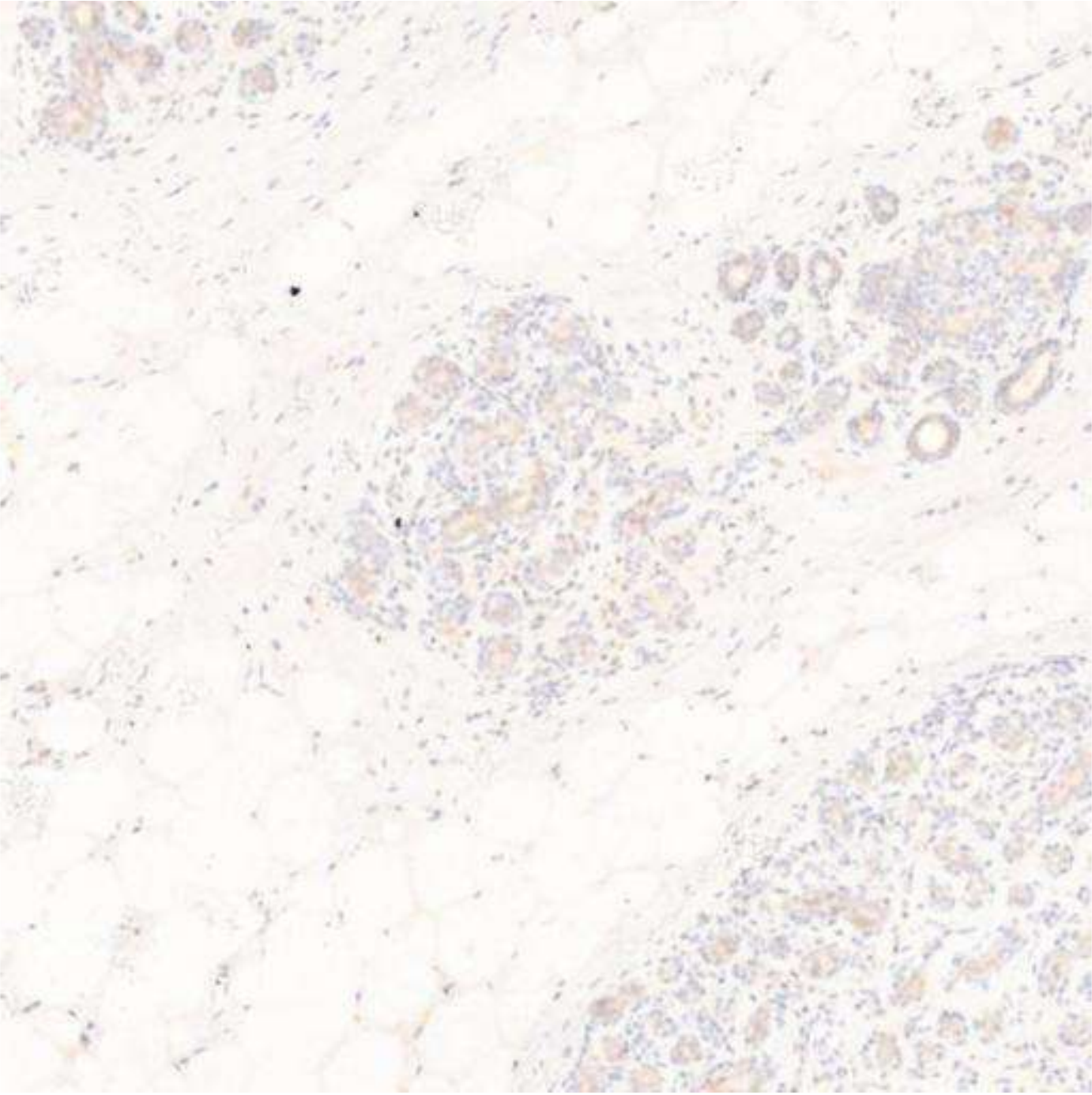}
    \caption{scale 2}
    \label{Octave2}
  \end{subfigure}
  \hfill
  \begin{subfigure}{0.24\linewidth}
    \includegraphics[width=1\linewidth]{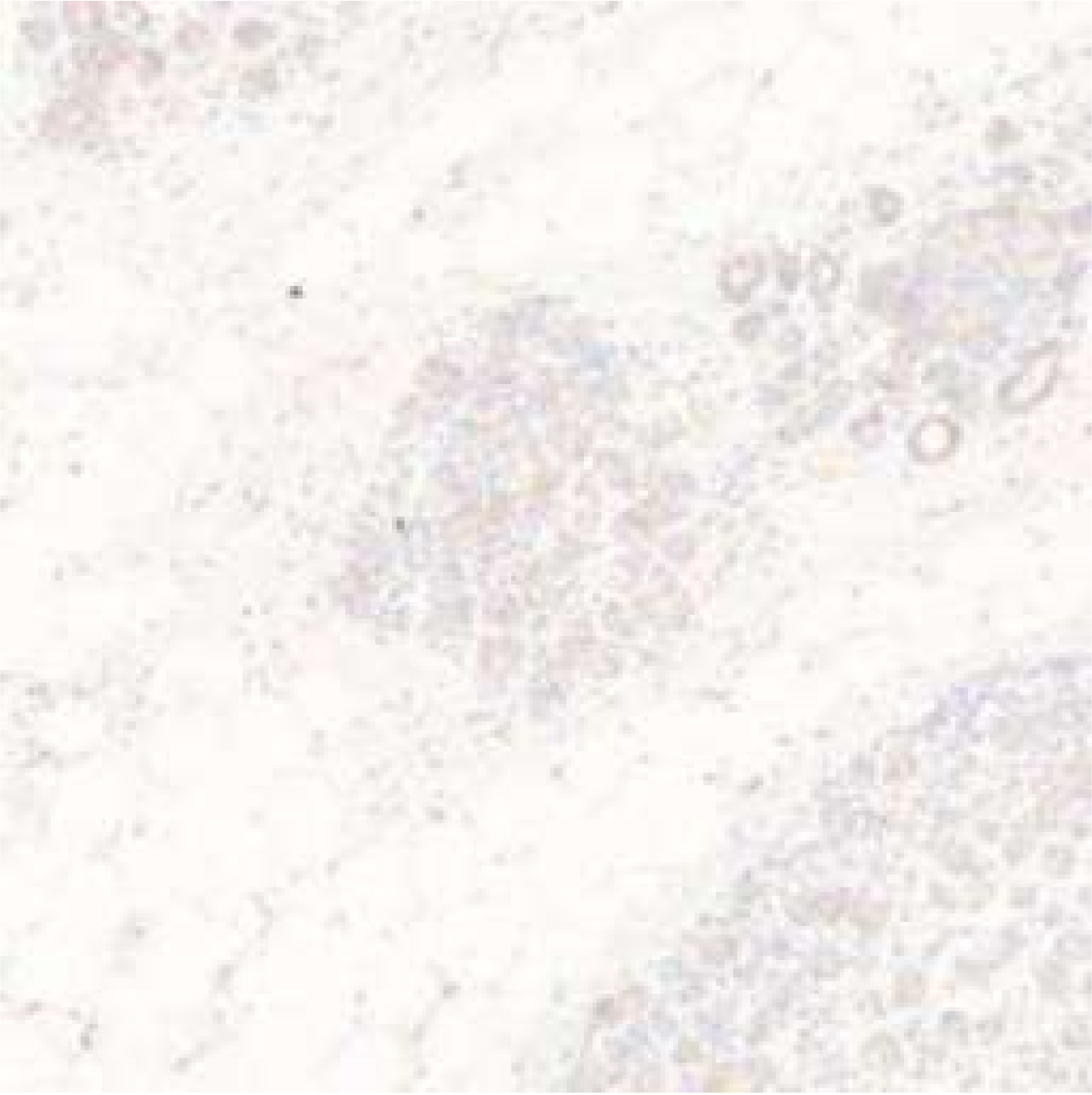}
    \caption{scale 3}
    \label{Octave3}
  \end{subfigure}
  \hfill
  \begin{subfigure}{0.24\linewidth}
    \includegraphics[width=1\linewidth]{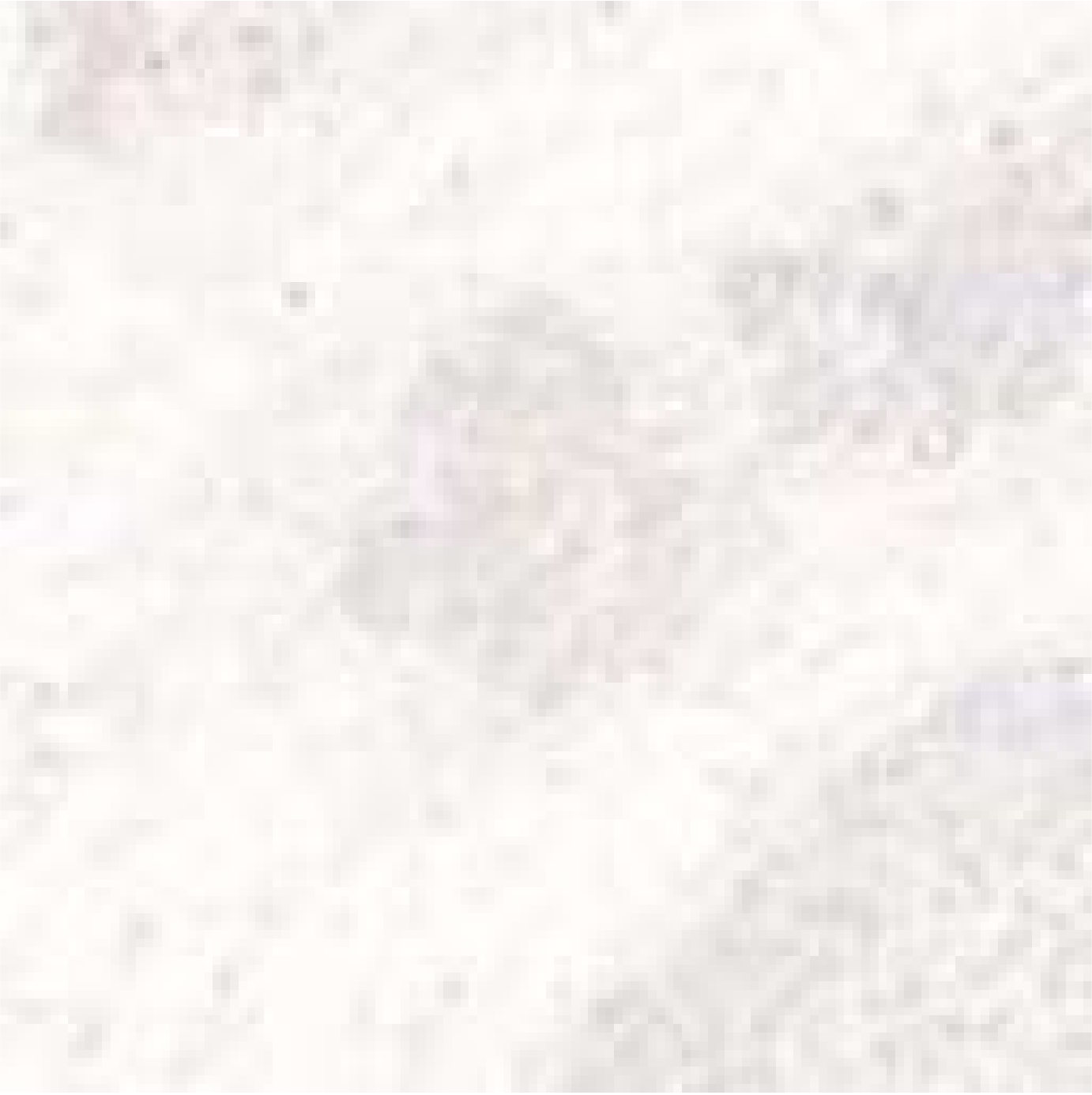}
    \caption{scale 4}
    \label{Octave4}
  \end{subfigure}
  \caption{Visualization of a sample image at different scales. Gaussian convolution makes the image gradually blurred.}
  \label{Output_of_octaves}
\end{figure}

The $L_1$ loss in pix2pix algorithm directly calculates the difference between the generated image and the ground truth, which is too restrictive on the generated image. For our BCI dataset, we need to weaken the constraints of $L_1$ loss, while aligning the generated image and ground truth at other scales. Inspired by scale-space theory\cite{lowe2004distinctive}, we will perform the same scale transformation on the generated image and ground truth. The scale transformation consists of two steps: 
1) Using a low-pass filter to smooth the image. 2)  Downsampling the smooth image.
Since the Gaussian kernel is the only linear kernel that realizes the image scale transformation, our low-pass filter uniformly uses the Gaussian kernel with a standard deviation of 1. With the progress of Gaussian filtering, the image becomes more and more blurred, and we reduce the resolution by downsampling to remove redundant pixels. For each resolution level (octave), multiple Gaussian convolutions are performed to achieve scale transformation.
Our pix2pix pyramid has several octaves, the first layer of each octave is obtained by down-sampling the last image of the previous octave; each octave has 5 layers and performs 4 Gaussian blurrings. For each output of octave, we define it as a scale (Fig.~\ref{Output_of_octaves}).
In our Gaussian Pyramid, we extract the first layer of images in each octave to calculate the loss. The loss for each scale is denoted as $S_i(i=1,2,3\cdots)$:
\begin{equation}
  S_i=\mathbb{E}_{x,y,z}[{\left\|F_i(y)-F_i(G(x,z))\right\|}_1]\text{,}
  \label{targetfunction}
\end{equation}
where $F_i$ and $G$ represent the Gaussian filtering operation and the generator, respectively. $x$, $y$ and $z$ represent the input image, the ground truth, and random noise, respectively. Even if we cannot make the generated image highly consistent with the ground truth in the first octave, we can still make the generated image close to the ground truth on a higher-dimensional scale. Our multi-scale loss is recorded as:
\begin{equation}
  L_{multi-scale}=\sum_{i}{\lambda_iS_i}\text{,}
  \label{targetfunction}
\end{equation}
\noindent where $\lambda_i$ represents the weight of scale $i$. Our adversarial loss is still consistent with pix2pix:
\begin{equation}
\begin{aligned}
  L_{cGAN}(G,D)=
&\mathbb{E}_{x,y}[logD(x,y)]+ \\
&\mathbb{E}_{x,z}[log(1-D(x,G(x,z)))]\text{,}\\
\end{aligned}
  \label{targetfunction}
\end{equation}
\noindent where generator $G$ tries to minimize this function while discriminator $D$ tries to maximize it. We still keep the $L_1$ loss in pix2pix in order to maintain the constraints on the original resolution:
\begin{equation}
 L_1=\mathbb{E}_{x,y,z}[{\left\|y-G(x,z)\right\|}_1]\text{.}
\end{equation}
At this point, our overall objective function is:
\begin{equation}
\begin{aligned}
 G^{*}=arg\min_{G}\max_{D}L_{cGAN}(G,D)+\lambda_1L_1+L_{multi-scale}\text{.}
\end{aligned}
  \label{targetfunction}
\end{equation}

\section{Experiments}
In this section, we will use several image translation algorithms to conduct experiments on our BCI dataset. Our experiment was performed on NVIDIA Tesla T4 16GB GPU.

\subsection{Implementation}

In the pix2pix algorithm, we tried two generator structures, unet256 and resnet-9blocks. Through experiments, we found that resnet-9blocks obviously has a better generation effect. Therefore, in our proposed method, we use the generator structure of resnet-9blocks as the baseline, while the discriminator structure uses the default patchGAN; the Gaussian kernel used is 3$\times$3 with a standard deviation of 1; the input images are not preprocessed; the batch size is set to 2; the optimizer used is Adam; the total number of training epochs is set to 100: the learning rate of first 50 epochs is set to 0.0002 and the learning rate of the remaining 50 epochs gradually drops to 0. In pix2pixHD, both the generator and the discriminator adopt the default settings, and the image preprocessing, the number of training epochs and the optimization strategy are consistent with those of pix2pix. In the cycleGAN algorithm, due to memory limitations, we randomly crop the image to 512$\times$512 resolution before training, and other settings are also consistent with pix2pix. 

\subsection{Metrics}

\begin{figure*}[hbt]
  \centering
  \begin{subfigure}{0.465\linewidth}
    \includegraphics[width=1\linewidth]{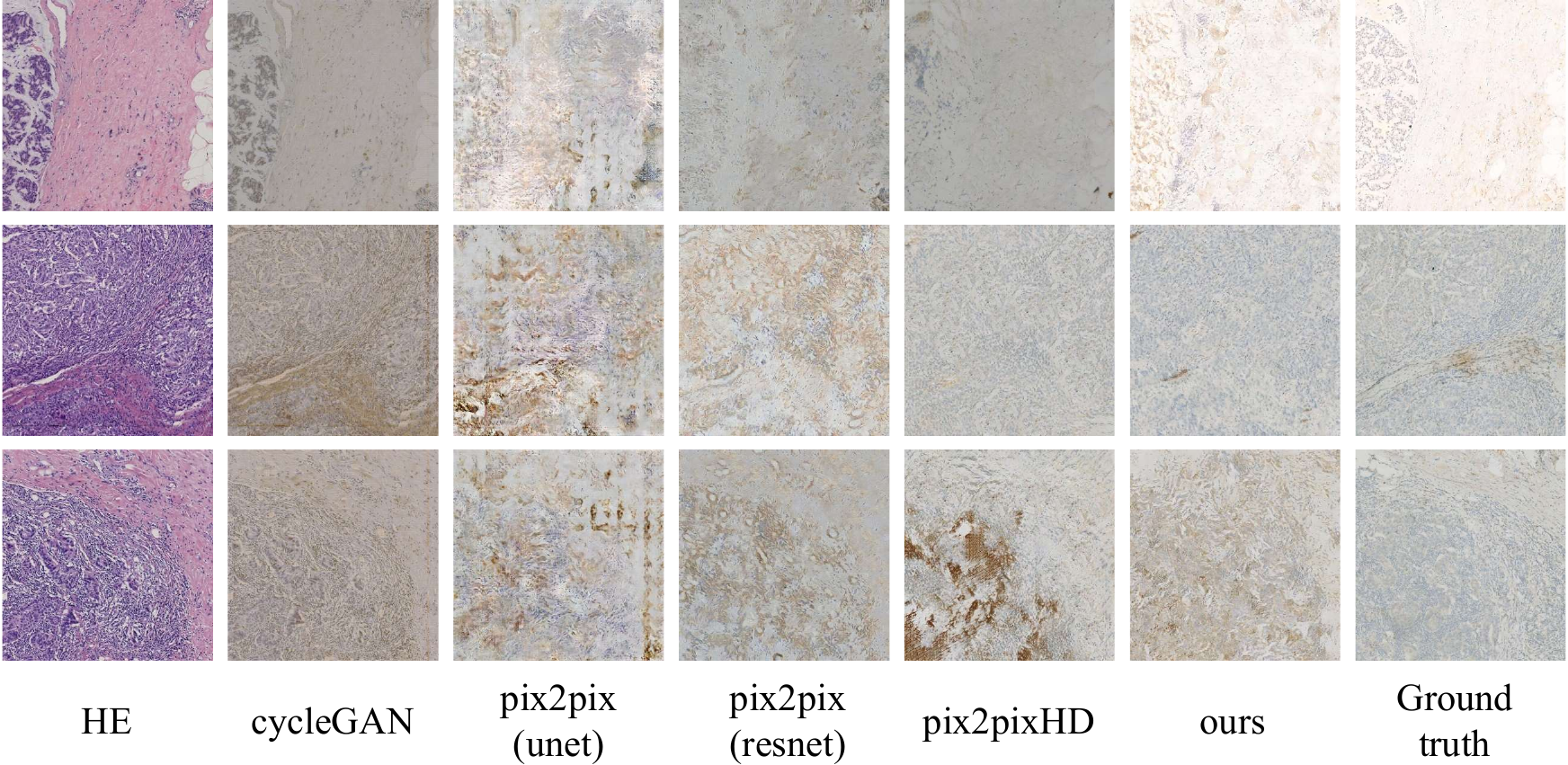}
    \caption{Visualization of different methods on IHC 0 images. The image we generated is very close to the ground truth.}
    \label{result_0}
  \end{subfigure}
  \hfill
  \begin{subfigure}{0.465\linewidth}
    \includegraphics[width=1\linewidth]{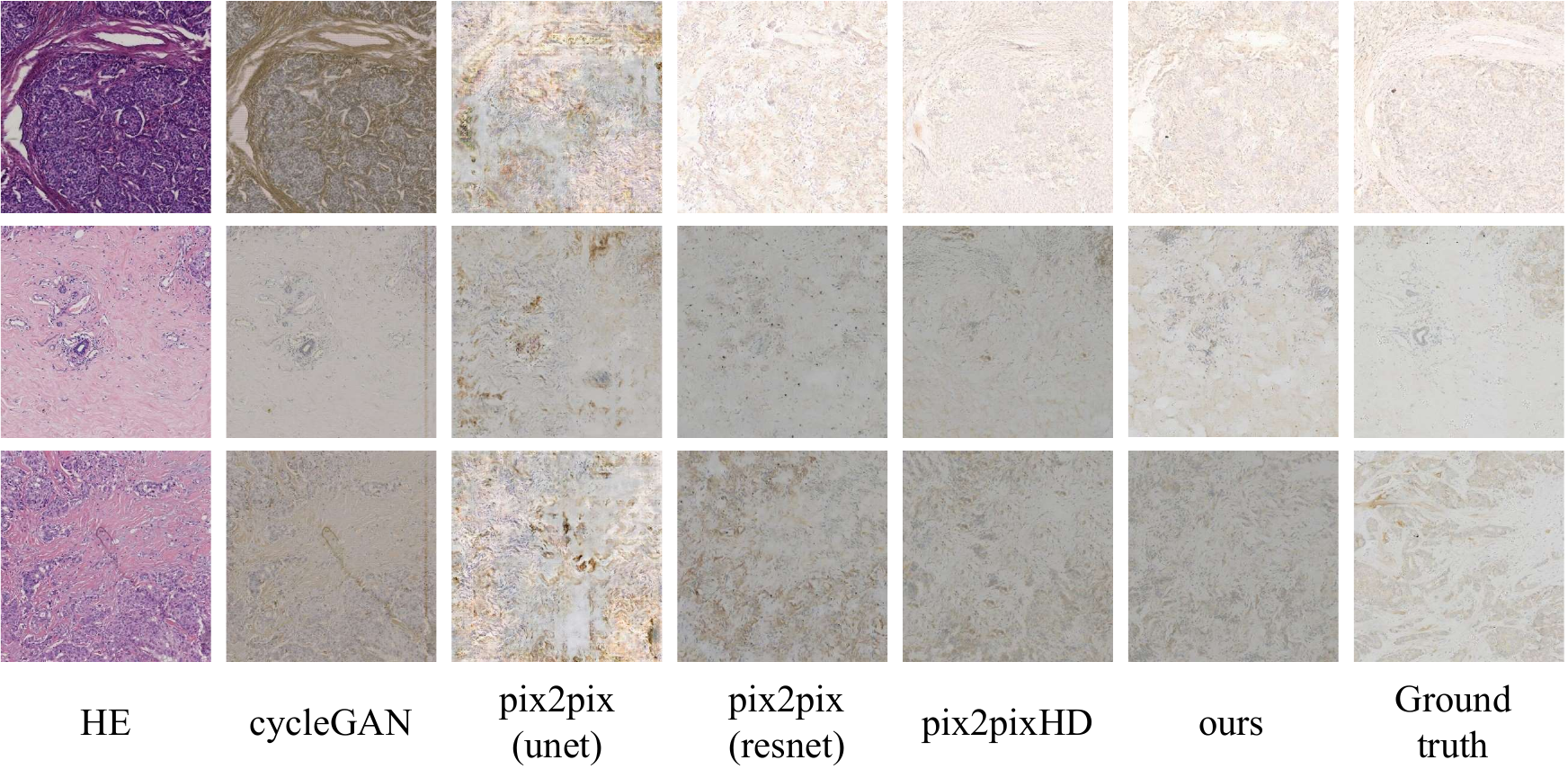}
    \caption{Visualization of different methods on IHC 1+ images. The image we generated is very close to the ground truth.}
    \label{result_1+}
  \end{subfigure}
  \hfill
  \begin{subfigure}{0.465\linewidth}
    \includegraphics[width=1\linewidth]{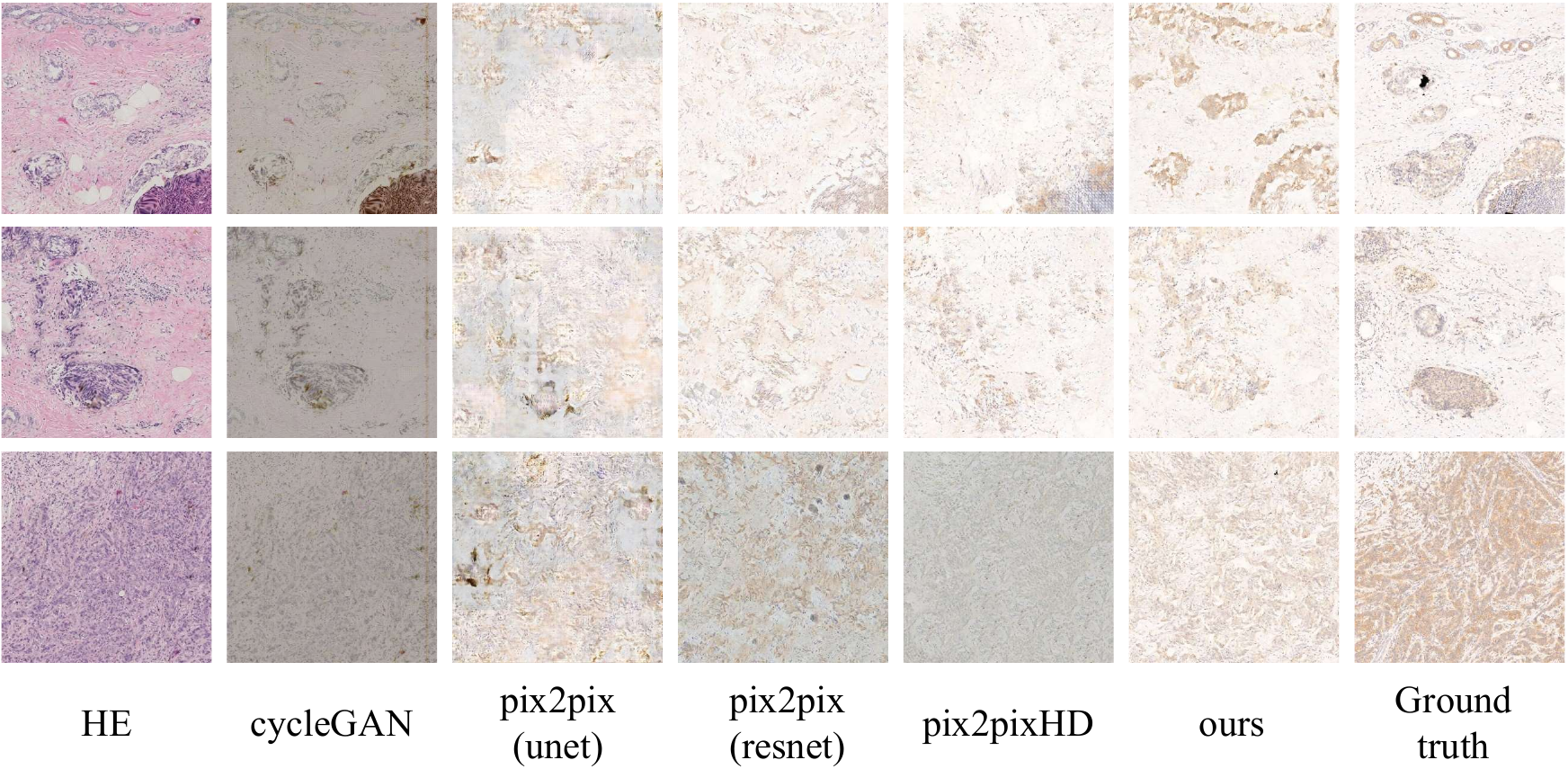}
    \caption{Visualization of different methods on IHC 2+ images. In general, our generated image has a lighter color than ground truth, however, it is still better than other methods.}
    \label{result_2+}
  \end{subfigure}
  \hfill
  \begin{subfigure}{0.465\linewidth}
    \includegraphics[width=1\linewidth]{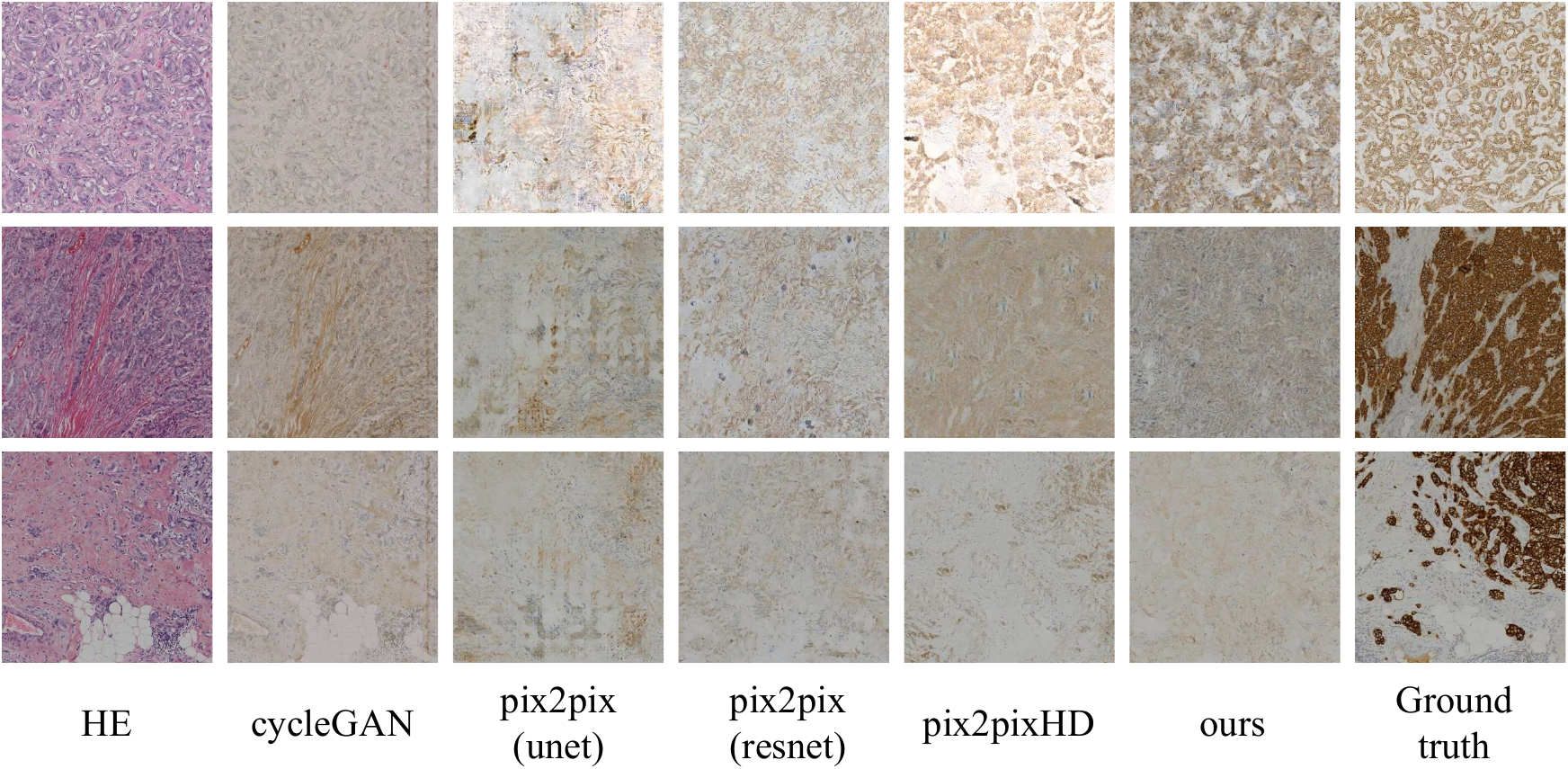}
    \caption{Visualization of different methods on IHC 3+ images. In this case, all methods are difficult to accurately identify the cancer area, which is a huge challenge.}
    \label{result_3+}
  \end{subfigure}
  \caption{Visualization of different methods on different HER2 expressions.}
  \label{overallresults}
\end{figure*}
We use Peak Signal to Noise Ratio (PSNR) and Structural Similarity (SSIM) as the evaluation indicators for the quality of the generated image. PSNR is based on the error between the corresponding pixels of two images and is the most widely used objective evaluation index. However, the evaluation result of PSNR may be different from the evaluation result of the Human Visual System (HVS). Therefore, we also use SSIM \cite{wang2004image,lowe1999object}, which comprehensively measures the differences in image brightness, contrast, and structure. This evaluation result is closer to the human visual system.


\subsection{Benchmark Results}

\begin{table}[htb]
\begin{center}{
\begin{tabular}{ccccccc}
\hline
                    Method  & PSNR(dB)  & SSIM   \\ \hline
cycleGAN              & 16.203    & 0.373  \\
pix2pix(unet generator)         & 18.654    & 0.419  \\ 
pix2pix(resnet generator)       & 19.328     & 0.440  \\ 
pix2pixHD             & 19.634    & 0.471 \\
ours                  & \textbf{21.160} & \textbf{0.477} \\    
\hline
\end{tabular}}
\end{center}
\caption{Comparison of PSNR and SSIM values of different methods on BCI dataset.} 
\label{BCI_metrics}
\end{table}

\begin{table}[htb]
\begin{center}{
\begin{tabular}{ccccccc}
\hline
                    Method  & PSNR(dB)  & SSIM   \\ \hline
cycleGAN              & 11.22    & 0.214  \\
pix2pix(unet generator)         & 10.769    & 0.176  \\ 
pix2pix(resnet generator)       & 12.082     & 0.207  \\ 
pix2pixHD             & 11.156    & 0.228 \\
ours                  & \textbf{12.191} & \textbf{0.278} \\    
\hline
\end{tabular}}
\end{center}
\caption{Comparison of PSNR and SSIM values of different methods on LLVIP dataset.} 
\label{LLVIP_metrics}
\end{table}

For the unsupervised method, we choose the most representative cycleGAN. It can be seen from the experimental results (Fig.~\ref{overallresults} and Table ~\ref{BCI_metrics}) that as an unsupervised image translation algorithm, cycleGAN cannot establish an accurate mapping from HE to IHC results. For these registered image pairs, it can only achieve ``style'' migration, but it is completely impossible to identify the cancer areas. As a representative algorithm of supervised image translation, pix2pix with a resnet generator can  basically stain the cancerous area. Its PSNR and SSIM indicators are significantly higher than cycleGAN, but the quality of the generated image is poor, pix2pix with unet generator is even worse. Besides, the staining effect of pix2pix generated images is quite different from the correct results of IHC, especially in the areas where HER2 is highly expressed. Pix2pixHD uses a two-stage generator structure and performs adversarial discriminating on multiple scales. Its high-resolution image generation quality is slightly better than pix2pix on the whole, therefore, it has higher PSNR and SSIM than pix2pix. However, in some areas with low HER2 expression, pix2pixHD may incorrectly generate dark browns.

The result of our method is better than pix2pix and pix2pixHD in terms of authenticity. In the identification of HER2 expression, our method is better in the case of low expression of HER2 (0/1+), the difference between the generated image and ground truth is slight (Fig.~\ref{overallresults}(a)(b)); when the expression level of HER2 is 2+, the image we generate will be lighter than ground truth, but the effect is still better than other methods (Fig.~\ref{overallresults}(c)); when HER2 is highly expressed (3+), our method is the same as other methods, unable to identify areas of high expression of HER2 (Fig.~\ref{overallresults}(d)), which is also a major issue that needs to be resolved in the future. It is still very challenging to establish an accurate mapping from HE to HER2 expression on our dataset. We still need to explore more effective methods to improve the accuracy of the translation.

On the LLVIP\footnote{Visit the link \url{https://bupt-ai-cz.github.io/LLVIP} for details of the LLVIP dataset} dataset, our method also achieves the best PSNR and SSIM (Table~\ref{LLVIP_metrics}), which proves that our method is not only suitable for the translation of pathological images but also has a certain versatility.

\subsection{Multi-scale Analysis}

\begin{table}[htb]
\begin{center}
{
\begin{tabular}{ccccccc}
\hline
             Configuration         & PSNR(dB)  & SSIM   \\ \hline
pix2pix      & 19.328    & 0.440  \\
pix2pix+S1 (ours)            & \textbf{21.160}   & \textbf{0.477}  \\ 
pix2pix+S1+S2 (ours)         & 21.033    & 0.469  \\ 
pix2pix+S1+S2+S3 (ours)      & 21.138    & 0.472 \\
\hline
\end{tabular}}
\end{center}
\caption{Multi-scale analysis on BCI dataset.} 
\label{BCI}
\end{table}

\begin{table}[htb]
\begin{center}
{
\begin{tabular}{ccccccc}
\hline
              Configuration        & PSNR(dB)  & SSIM   \\ \hline
pix2pix      & 12.082    & 0.207  \\
pix2pix+S1 (ours)            & {12.189}   & \textbf{0.279}  \\ 
pix2pix+S1+S2 (ours)         & 12.173    & 0.277  \\ 
pix2pix+S1+S2+S3 (ours)      & \textbf{12.191} & {0.278} \\
\hline
\end{tabular}}
\end{center}
\caption{Multi-scale analysis on LLVIP dataset.} 
\label{LLVIP}
\end{table}

Our pyramid pix2pix has the flexibility to change the number of pyramid layers to accommodate different datasets. On our BCI dataset, we tried the gains of different pyramid levels. Table~\ref{BCI} shows that the model with a two-layer pyramid structure (pix2pix+S1) achieves the highest PSNR and SSIM, which demonstrates that it is more reasonable to constrain the generated images and ground truth on the second scale. By optimizing the loss function of scale two (S1), the model effect can be greatly improved. On LLVIP dataset, a four-layer pyramid pix2pix can achieve the approximate effect of a two-layer model (Table~\ref{LLVIP}), which shows that the constraint of the high level also improves the generation effect compared to pix2pix.
\subsection{Subjective Validation}

In addition to objective metrics, we also invited two pathologists to diagnose HER2 expression of the generated images. To avoid the influence of subjective factors, we randomly selected 40 real-generated IHC pairs, and shuffled the order of these 80 images. A generated IHC image is considered accurate if the generated image and its corresponding real image are diagnosed at the same level. The accuracy of these 40 generated images is shown in Table~\ref{subjective}. The results show that there is still a long way to go for current methods before clinical application, which also proves the importance of the BCI dataset in further research.

\begin{table}[htb]
\begin{center}
{
\begin{tabular}{ccccccc}
\hline
                     & pathologist1  & pathologist2   \\ \hline
Accuracy(\%)      & 37.5    & 40.0  \\
\hline
\end{tabular}}
\end{center}
\caption{Accuracy of generated IHC images.} 
\label{subjective}
\end{table}
\section{Conclusion}

In this paper, we propose BCI, a new dataset in the field of pathology images for the translation of HE stained breast tissue section to its IHC results. This task puts forward new requirements for image translation algorithms, which is to accurately identify the expression area and expression level of HER2 while ensuring the authenticity of the generated image. In addition, we also propose pyramid pix2pix, an image-to-image translation model suitable for registered image pairs. 

It is still very challenging to establish an accurate mapping from HE to IHC results. There is still a need for more effective methods to improve the accuracy of the translation. In addition, in the future, we will explore the difference in HER2 evaluation between synthetic IHC images and real IHC images. Then we will further study the possibility of formulating accurate clinical treatment plans for breast cancer using synthetic IHC images.
\section*{Acknowledgement}
This work was supported in part by the National Natural Science Foundation of China under Grant 62176167, and in part by the BUPT innovation and entrepreneurship support program under Grant 2022-YC-T046.

{\small
\bibliographystyle{ieee_fullname}
\bibliography{ReviewTemplate}
}

\end{document}